\documentclass[galaxies,review,accept,pdftex,moreauthors]{Definitions/mdpi} 

\firstpage{1} 
\pubvolume{12}
\issuenum{1}
\articlenumber{46}
\pubyear{2024}
\copyrightyear{2024}
\externaleditor{Academic Editor: Margo Aller}
\datereceived{ 30 June 2024} 
\daterevised{ 23 July 2024} 
\dateaccepted{ 3 August 2024} 
\datepublished{  7 August 2024} 
\hreflink{https://doi.org/10.3390/\\galaxies12040046} 

\usepackage{xspace}

\def\apss{Astrophys. Space Sci.}   
\def\aap{Astron. Astrophys.}                
\def\aj{Astron. J.}

\def\apj{Astrophys. J.}   
\def\apjl{Astrophys. J. Lett.}   
\def\aplett{Astrophys. Lett.}

\def\mnras{Mon. Not. R. Astron. Soc.}

\def\nat{Nature}

\def\pasj{Publ. Astron. Soc. Jpn.}

\def\prd{Phys. Rev. D}

\newcommand*{\vect}[1]{\mbox{\boldmath ${#1}$}}

\newcommand {\her}{\mbox{Her\,X-1}\xspace}
\newcommand {\rx}{\mbox{RX\,J0440.9+4431}\xspace}
\newcommand {\lsv}{\mbox{LS\,V\,+44~17}\xspace}

\newcommand {\swiftp}{\mbox{Swift\,J0243.6+6124}\xspace}

\newcommand\degr{\ensuremath{^\circ}}
\newcommand\ixpe{\textit{IXPE}\xspace}

\Title{X-ray Polarimetry of X-ray Pulsars}

\TitleCitation{X-ray Polarimetry of X-ray Pulsars}

\Author{Juri 
 Poutanen *\orcidA{}, Sergey S. Tsygankov 
 \orcidB{} and Sofia V. Forsblom \orcidC{}}

\AuthorNames{Juri Poutanen, Sergey S. Tsygankov and Sofia V. Forsblom}

\AuthorCitation{Poutanen, J.; Tsygankov, S.S.; Forsblom, S.V.}

\address[1]{%
 Department of Physics and Astronomy,  FI-20014  University of Turku,  Finland; \\ 
 sergey.tsygankov@utu.fi~(S.S.T.); sofia.v.forsblom@utu.fi (S.V.F.)} 

\corres{\hspace{-0.82em} Correspondence: juri.poutanen@utu.fi}

\abstract{Radiation from X-ray pulsars (XRPs) was expected to be strongly linearly polarized owing to a large difference in their ordinary and extraordinary mode opacities. 
The launch of \ixpe  allowed us to check this prediction. 
\ixpe observed a dozen X-ray pulsars, discovering pulse-phase dependent variation of the polarization degree (PD) and polarization angle (PA). 
Although the PD showed rather erratic profiles resembling flux pulse dependence, the PA in most cases showed smooth variations consistent with the rotating vector model (RVM), which can be interpreted as a combined effect of vacuum birefringence and dipole magnetic field structure at a polarization-limiting (adiabatic) radius. 
Application of the RVM allowed us to determine XRP geometry and to confirm the free precession of the NS in \mbox{Her X-1}. 
Deviations from RVM in two bright transients led to the discovery of an unpulsed polarized emission likely produced by scattering off the accretion disk wind.
}


\begin{document}


\section{Introduction}

Classical X-ray pulsars (XRPs) are neutron stars (NSs) found in binary systems, accreting material from companion stars via an accretion disk or stellar wind (for a review, see~\citep{MushtukovTsygankov2024}). 
The ultra-strong magnetic field of the NSs of the order of 10$^{12\text{--}13}$~G is able to  significantly alter the accretion flow's structure at distances of hundreds of NS radii. 
At this distance, known as the magnetospheric radius, the~magnetic pressure balances the ram pressure in the disk. 
Consequently, closer to the NS, ionized gas is funneled along magnetic field lines to the magnetic poles, where its kinetic energy is released in the form of~X-rays.

The geometrical structure of this emission region and the corresponding beam pattern of an XRP depend strongly on the mass accretion rate. Particularly, it was demonstrated theoretically that there exists a so-called critical luminosity defining two major regimes of accretion \citep{Basko76, Becker12,Mushtukov15}. 
Below this threshold, emission originates from the hot spots at the NS's surface, while above it, the~radiation pressure is able to stop the infalling matter above the star surface, and an accretion column begins to grow, thus drastically changing the structure of the emitting region. 
The absolute value of the critical luminosity is a non-monotonic function of the magnetic field strength and has a typical value of 10$^{37}$ erg~s$^{-1}$ \cite{Mushtukov15}.

The primary observables from XRPs, including energy spectra and light curves, contain information about the geometrical configuration of the emitting regions at the NS surface. 
However, this information is heavily degenerate with very complex and uncertain physics of emission mechanisms and radiation transfer under the conditions of extremely strong magnetic fields. 
Therefore, an~independent tool sensitive to the geometry is required to address the longstanding challenge of determining the configurations of the emitting regions in~XRPs.

Existing theoretical models predicted XRPs to be among the most polarized sources in the X-ray sky. 
This is due to the birefringence of the medium in a strong magnetic field. 
Specifically, photon propagation through such a medium can be considered in terms of two normal polarization modes: the so-called ordinary `O' and extraordinary `X' photons~\cite{Gnedin74,Gnedin78}. 
Due to the different orientations of the photons' electric field vector oscillation with respect to the plane defined by the directions of the magnetic field and photon momentum, the~cross section of interaction between these photons and matter is very different for the two modes. 
Namely, below~the cyclotron frequency, the~Compton scattering cross section for the O-mode radiation is much lower than that for the X-mode (see,, for example,~\citep{Harding06,Mushtukov16}). 
Many authors have addressed the polarization properties of XRPs \citep{1981ApJ...251..278N,1981ApJ...251..288N,1982Ap&SS..86..249K,1985ApJ...298..147M,1985ApJ...299..138M,1987PASJ...39..781K,Meszaros88}, often predicting a very high polarization degree (PD), reaching nearly 100\%. 
The most recent models cover the low \citep{Mushtukov21,2021A&A...651A..12S} and the high \citep{Caiazzo21} mass accretion rates with a thin slab and accretion column geometries, respectively, also predicting a substantial PD in excess of 30\% in the standard X-ray range 2--10~keV. 
A relatively low polarization degree (PD) remains possible at low mass accretion rates due to the inverse temperature profile in the atmosphere of an accreting NS \citep{2019MNRAS.483..599G,Mushtukov21}. 
The verification of different model predictions requires sensitive polarimetric~observations.

\begin{figure}[H] 
\includegraphics[width=0.8\linewidth]{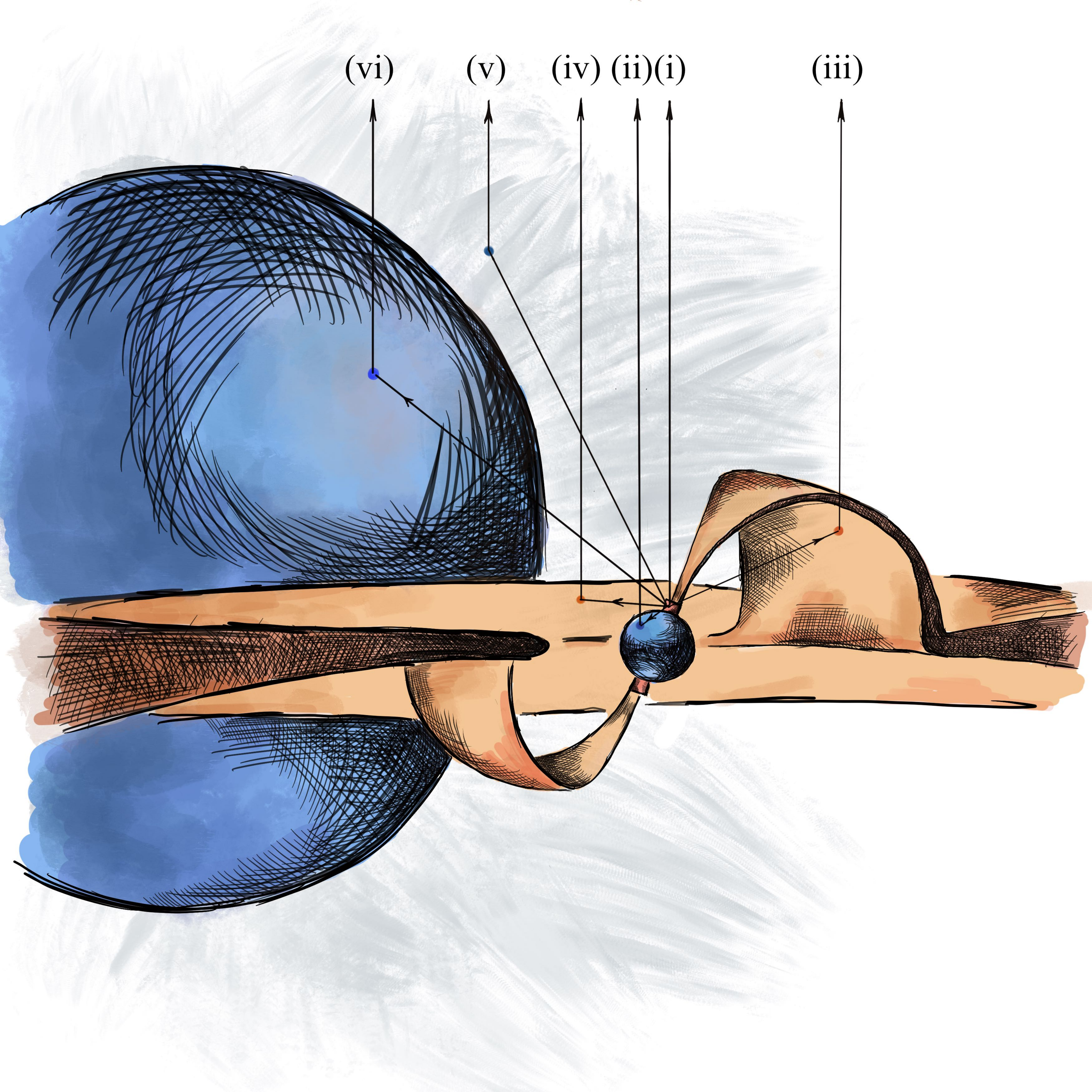}
\caption{Possible sources of polarized emission in XRPs: 
(i) intrinsic polarization from the hotspot, (ii) reflection from the NS surface, (iii) reflection from the accretion curtain, (iv) reflection from the accretion disk (and accretion disk wind), (v) scattering off the stellar wind, and (vi) reflection off the optical companion. 
From~\cite{Mushtukov23}.}
\label{fig:illustration} 
\end{figure}

In addition to the emission produced at the magnetic poles, one expects contributions to the polarized emission also from the reflection of that  radiation off the NS surface, accretion disk, accretion disk wind, accretion curtain, stellar wind, and~stellar surface (see Figure~\ref{fig:illustration}). 
Some of these components are expected to be pulsating with the pulsar frequency, some may be pulse-phase independent, and~some may depend on the orbital phase. 
Separating the contribution of these components is a difficult task, and X-ray polarimetry may provide an avenue for the~solution.

A handful of attempts have been made to detect the polarization of XRPs. 
Already in 1975, {OSO-8} observed \mbox{Cen X-3} and \mbox{Her X-1} to search for X-ray polarization  \citep{1979ApJ...232..248S} with no significant polarization detected either in pulse phase-averaged or phase-resolved data. 
The balloon-borne hard X-ray polarimeter X-Calibur operating in the 15--35 keV energy range observed GX~301$-$2, resulting in no significant detection of polarization    \citep{2020ApJ...891...70A}.
The successful launch in December 2021 of the \textit{Imaging X-ray polarimetry Explorer} (\ixpe; \cite{Weisskopf2022}), which has a sensitivity of at least two orders of magnitude higher than \mbox{OSO-8} \cite{2021AJ....162..208S}, allowed us to obtain significant detection of polarization in a dozen XRPs and to test theoretical models. 
Here, we review the results of the \ixpe observations of XRPs during the first two years of operations in~orbit.

\section{Rotating Vector~Model}
\label{sec:rvm}

The rotating vector model (RVM) was introduced in the 1960s  \citep{Radhakrishnan69,Komesaroff70} to describe the polarization from radio pulsars. 
In this model, the~evolution of the PA with pulsar phase is related  to the projection of the magnetic dipole on the plane of the sky.  
The same model was suggested by \citet{Meszaros88} to be used for modeling the X-ray polarization of XRPs. 
The main reason why the RVM should work for strongly magnetized NSs is that because of vacuum birefringence \citep{Gnedin78,Pavlov79}, photon polarization direction is adjusted to the local magnetic field geometry  until  the adiabatic radius is reached (e.g., \citep{Heyl00,Heyl02PRD,Taverna15}). 
This radius is estimated to be $\sim$20 NS radii for photons in the \ixpe\ range and the typical surface magnetic field strength detected in XRPs \citep{Heyl18,Taverna24}: 
\begin{linenomath}
\begin{equation}
R_{\rm ad}\sim 2\times 10^7\,\left(\frac{E}{4\,{\rm keV}}\right)^{1/5}\,
\left(\frac{B}{10^{12}\,\mbox{G}}\right)^{2/5}
\left(\frac{R_{\rm NS}}{12\,\mbox{km}}\right)^{6/5}\,\mbox{cm}.  
\end{equation} 
\end{linenomath}
At such distances, the~field is expected to be dominated by the dipole component. 
We expect, therefore, to~measure polarization either parallel or perpendicular to the instantaneous projection of the  magnetic dipole onto the plane of the sky depending on the intrinsic polarization mode. 
The phase variation of the PA is thus a purely geometrical effect and does not depend on the behavior of the PD or~flux.

\begin{figure}[H] 
\includegraphics[width=5 cm]{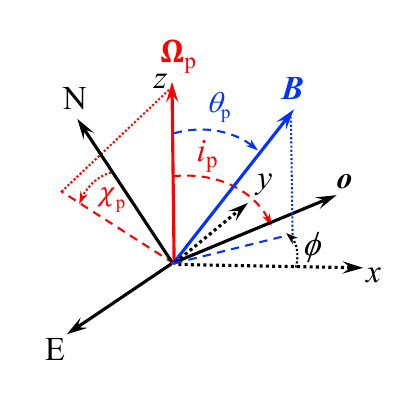}
\caption{Geometry of the pulsar and main parameters of the RVM. 
The pulsar angular momentum $\vect{\Omega}_{\rm p}$ makes an angle $i_{\rm p}$ with respect to the observer direction $\vect{o}$. 
The angle $\theta_{\rm p}$ is the magnetic obliquity, i.e.,~the angle between magnetic dipole $\vect{B}$ and the rotation axis. 
Pulsar phase $\phi$ is the azimuthal angle of vector $\vect{B}$ in the plane $(x,y)$ perpendicular to  $\vect{\Omega}_{\rm p}$ measured from the projection of $\vect{o}$. 
The pulsar position angle $\chi_{\rm p}$ is the angle measured counterclockwise between the direction to the north (N) and the projection of $\vect{\Omega}_{\rm p}$ on the plane of the sky (N-E).}\label{fig:geometry} 
\end{figure}

If radiation is dominated by the O-mode with the oscillation direction in the plane of the magnetic field and photon momentum, the~PA $\chi$ measured from the projection of the spin axis on the sky in the counterclockwise direction is given by \citep{Poutanen2020}: 
\begin{linenomath}
\begin{equation} \label{eq:pa_rvm}
\tan (\chi-\chi_{\rm p})=\frac{-\sin \theta_{\rm p}\ \sin \phi}
{\sin i_{\rm p} \cos \theta_{\rm p}  -  \cos i_{\rm p} \sin \theta_{\rm p}  \cos \phi } ,
\end{equation} 
\end{linenomath} 
where $\chi_{\rm p}$ is the position angle (measured from north to east) of the pulsar angular momentum, $i_{\rm p}$ is the inclination of the pulsar spin with respect to the line of sight, and $\theta_{\rm p}$ is the magnetic obliquity, i.e.,~the angle between the magnetic dipole and the spin axis (see Figure~\ref{fig:geometry} for geometry).

For radiation in the X-mode, the~PA $\chi$ is rotated by 90\degr\ with respect to the O-mode. 
We note that in slow XRPs, the general relativistic effects such as light bending do not affect the expression for the PA because~photon trajectories are planar in the Schwarzschild metric. 
For NSs rotating at millisecond periods, the~special relativity effects need to be accounted for, and~also the general relativistic effects become important~\cite{Poutanen2020,Loktev20}.

\begin{figure}[H]

\begin{adjustwidth}{-\extralength}{0cm}
\centering
 \includegraphics[width=7.8cm]{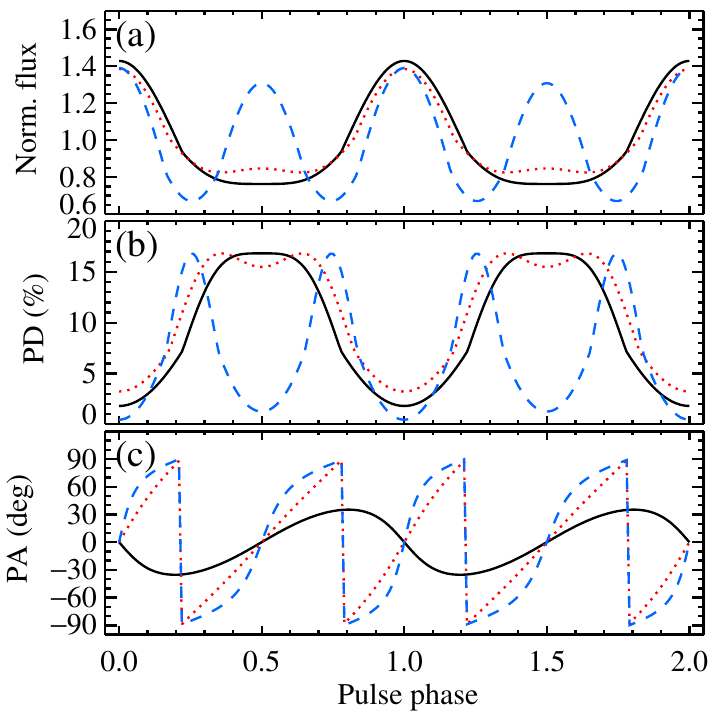}
\hspace*{0.3cm}
\includegraphics[width=8.0cm]{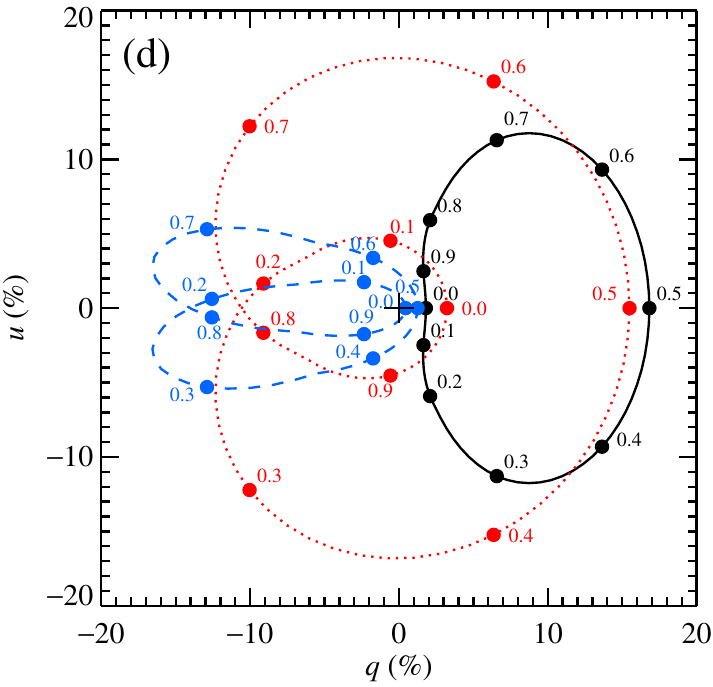}
\end{adjustwidth}
\caption{Examples of the RVM with the emission pattern and polarization given by Equations~(\ref{eq:Ialpha}). 
(\textbf{a}) Normalized flux, (\textbf{b}) PD, and~(\textbf{c}) PA as functions of pulse phase for various pairs of parameters $(i_{\rm p},\theta_{\rm p})$. 
The case of (60\degr,\,30\degr) is shown with black solid lines,  (30\degr,\,70\degr) with red dotted lines, and~(70\degr,\,85\degr) with blue dashed lines. 
(\textbf{d}) Evolution of normalized Stokes parameters $(q,u)$ for the same three sets of pairs. The~numbers mark the pulse phase. 
We use here the following parameters: the NS mass $M=1.4M_{\odot}$ and radius of 12~km, $a=1$, $b=0.5$, $P_{\max}=30\%$, and~ $\chi_{\rm p}=0$. 
\label{fig:rvm}}
\end{figure}

Let us now consider a simple toy model for XRP polarization. 
We consider emission from two point-like antipodal hotspots at the NS  surface. 
We assume the intensity and the PD of radiation escaping from the hotspot to vary with the zenith angle $\alpha$ as 
\begin{linenomath}
\begin{equation} \label{eq:Ialpha}
 I(\alpha)= I_0\ (1+a\cos\alpha), \quad 
P  = P_{\max}(1-\cos\alpha)/(1+b\cos\alpha) .  
\end{equation} 
\end{linenomath} 
Positive $a$ corresponds to the pencil-like emission pattern, while the negative $a$ to the fan-like pattern. 
The expression for the PD implies that polarization is zero along the normal to the surface, $\alpha=0$, because~of the axial symmetry and monotonically growing with the zenith angle.  
The angle between the magnetic dipole (normal to the primary spot) and the direction to the observer $\psi$ varies with the pulsar phase $\phi$ as:
\begin{linenomath}
\begin{equation}
\cos\psi_{\rm p}=\cos i_{\rm p}\cos \theta_{\rm p}+\sin i_{\rm p}\sin \theta_{\rm p}\cos\phi. 
\end{equation} 
\end{linenomath} 
For the secondary spot in the southern hemisphere, $\cos\psi_{\rm s}=-\cos\psi_{\rm p}$.
For a slowly rotating NS, we can use the Schwarzschild metric to describe gravitational light bending. 
For calculations, we use an accurate analytical expression for the relation $\alpha(\psi)$ given by Equation~(2) in~\cite{Poutanen2020b}. 
The observed flux is then $F \propto {\cal D} \ I(\alpha) \cos\alpha$, 
with the lensing factor  ${\cal D}\propto d\cos\alpha/d\cos\psi$   given by Equation~(17) in~\cite{Poutanen2020b}. 
The visibility condition for every spot reads $\cos\alpha>0$. 
The Stokes parameters for each spot are given by the products $Q_i=F_i\,P_i\,\cos2\chi$ and $U_i=F_i\,P_i\,\sin2\chi$. 
Summing up the Stokes vectors for two spots $F=F_1+F_2$, $Q=Q_1+Q_2$, and~$U=U_1+U_2$, we obtain the total Stokes vector describing the linearly polarized light observed from the XRP.   
The normalized Stokes parameters can be defined as $q=Q/F$ and $u=U/F$.
To demonstrate what kind of polarization is expected in this model and to develop physical intuition, we compute three cases reflecting different geometries. 
The pulse phase dependencies of the normalized flux, the~PD, and~PA together with the trajectories and XRP left on the $(q,u)$ plane are shown in Figure~\ref{fig:rvm}.

The first thing to notice is the apparent anticorrelation between the flux and PD, which is a result of assuming a pencil-beam emission with $a>0$. 
For $a<0$, one expects a correlation. 
In the first case (black solid lines), $(i_{\rm p},\theta_{\rm p})$=(60\degr,\,30\degr), the~observer is looking at the magnetic dipole from a side, and both hotspots are visible all the time. 
The PA shows an S-shape profile. 
We note that the clockwise rotation of the pulsar (i.e., replacing $i_{\rm p} \rightarrow 180\degr-i_{\rm p}$) produces PA of an opposite sign (and shifted by half a period), and therefore can be easily distinguished from the counterclockwise rotation. 
The $(q,u)$ trajectory is a `kidney'-like shape outside of the origin, with~the exact shape, of~course, depending on the angular distributions of $I(\alpha)$ and $\mbox{PD}(\alpha)$. 
In the second case (red dotted lines), $(i_{\rm p},\theta_{\rm p})$=(30\degr,\,70\degr), the~observer sees the position angle of the dipole making a $360\degr$ rotation during the period, which is reflected in the sawtooth behavior of the phase dependence of the PA with a nearly constant derivative dPA/d$\phi$. 
For the counterclockwise rotation of the XRP $i_{\rm p}<90\degr$, the~PA grows with the phase, while for $i_{\rm p}>90\degr$, the~PA decreases. 
The source leaves a pretzel-like trace in the $(q,u)$ plane with two loops going around the origin. 
The third considered case (blue dashed lines), $(i_{\rm p},\theta_{\rm p})$=(70\degr,\,85\degr), corresponds to a nearly orthogonal rotator with the observer at a high inclination. 
We see 2-peak pulses corresponding to the two hotspots coming close to the line of sight with a strong variations in the PD reaching maxima at minima fluxes at phases 0.25 and 0.75. 
The PA still shows a sawtooth appearance but~with a strongly varying derivative dPA/d$\phi$. 
The trajectory at the $(q,u)$ plane has a rather complicated moth-like shape with two close approaches to the origin at phases 0 and 0.5. 
A non-zero pulsar spin position angle $\chi_{\rm p}$ results in a corresponding shift of the PA and rotation of the trajectories in the $(q,u)$ plane by the angle~$2\chi_{\rm p}$.

\section{X-ray Pulsars Observed by \ixpe}

During the first two years of operations in orbit, \ixpe observed eleven XRPs, with~several of them a few times in different states (see Table~\ref{tab:sample}). 
The  results are presented in a series of papers~\cite{Doroshenko22,Marshall22,Tsygankov22,Forsblom2023,Malacaria23,Tsygankov2023,Mushtukov23,Doroshenko23,Suleimanov2023,Heyl24,Zhao2024,Poutanen24,Forsblom24}. 
As discussed previously, the~high PD from magnetized NS is expected due to the large difference in cross sections between the two polarization modes at energies below the cyclotron resonance. 
As can be seen from the table, the~working energy range of \ixpe (2--8~keV) is significantly below the cyclotron energy for all XRPs in our sample, fulfilling the conditions for a high~PD.
\begin{table}[H]
\caption{Parameters of XRPs observed by \ixpe.}
\begin{adjustwidth}{-\extralength}{0cm}
\begin{tabularx}{\fulllength}{CCCCCcc}
\toprule
\textbf{Name}	&  \textbf{Spin Period \textsuperscript{a,b}}	& \textbf{Orbital Period~\textsuperscript{a,b}}     & \textbf{Distance~\textsuperscript{a,b}}      & \textbf{Luminosity~\textsuperscript{c}}     & \textbf{CRSF~\textsuperscript{d}}    & \textbf{\ixpe} \\
 &  \textbf{[s]}	& \textbf{[d]}   & \textbf{[kpc]}      & \textbf{[erg\,s$^{-1}$]}\  & \textbf{[keV]}     &  \textbf{Reference} \\
			\midrule
\mbox{Cen X-3} 	& 4.8			& 2.09			& 6.07			& $1.9\times10^{37}$		& 28        & \cite{Tsygankov22} \\
\mbox{Her X-1}     & 1.24			& 1.7			& 7.09			& $\sim$$3\times10^{37}$			& 37        & \cite{Doroshenko22,Heyl24} \\
\mbox{4U 1626$-$67}   & 7.7			& 0.02875			& 15.08			& $6.4\times10^{36}$			& 37, 61?   
    & \cite{Marshall22}  \\
\mbox{Vela X-1}   & 283			& 8.96			& 1.87			& $3.8\times10^{35}$			& 25, 53        & \cite{Forsblom2023}   \\
\mbox{GRO J1008$-$57}    & 93.5			& 249.5			& 3.21			& $(0.6\text{--}1.6)\times10^{36}$			& 78        & \cite{Tsygankov2023}   \\
\mbox{EXO 2030+375}   & 41.31			& 46.02			& 2.08			& $1.3\times10^{36}$			& 36/63? & \cite{Malacaria23}   \\
\mbox{X Persei}   & 837.67			& 250.3			& 0.63			& $1.2\times10^{34}$			& 29        & \cite{Mushtukov23}   \\
\mbox{GX 301$-$2}   & 696.0			& 41.59			& 3.54			& $1.3\times10^{36}$			& 37/50        & \cite{Suleimanov2023}   \\
\mbox{LS V +44 17}    & 202.5			& 155.0			& 2.29			& $\leq$$4\times10^{37}$			& 32        & \cite{Doroshenko23}    \\
\mbox{Swift~J0243.6+6124}   & 9.87			& 28.3			& 5.2			& $(0.6\text{--}2.4)\times10^{37}$			& 146        & \cite{Poutanen24}    \\
\mbox{SMC X-1}   & 0.717			& 3.892			& 61			& $2\times10^{38}$ & -- & \cite{Forsblom24} \\
\bottomrule
\end{tabularx}
\end{adjustwidth}
\noindent{\footnotesize{\textsuperscript{a} XRBcats: Galactic High Mass X-ray Binary Catalog~\cite{2023-HMXBcat}. \textsuperscript{b} XRBcats: Galactic low-mass X-ray binary catalog~\cite{2023-LMXBcat}. \textsuperscript{c} X-ray luminosity during \ixpe observations. \textsuperscript{d} Energy of the cyclotron resonance spectral feature from~\cite{2019-crsf}. Question mark identifies uncertain detections. }} 
 \label{tab:sample}
\end{table}
Observations of XRPs with \ixpe started with the first two sources of this class discovered at the dawn of X-ray astronomy, \mbox{Cen X-3} and \mbox{Her X-1}.
Both sources have a nearly critical luminosity of around 10$^{37}$~erg~s$^{-1}$. 
First and the most striking discovery made for both sources is a rather low PD, much below all theoretical predictions. 
Particularly, for~\mbox{Cen X-3}, the PD was found to be only $5.8\pm0.3\%$ in the phase-averaged data \citep{Tsygankov22}. 
Here, it is worth mentioning that due to the rotation of the NS, one might expect a reduction in the PD in the phase-averaged signal. 
Therefore, for~all \ixpe papers on XRPs, a lot of attention was given to the accurate timing solution required for the proper phase-resolved polarimetric~analysis.

\begin{figure}[H]
\begin{adjustwidth}{-\extralength}{0cm}
\centering
\includegraphics[width=8.4cm]{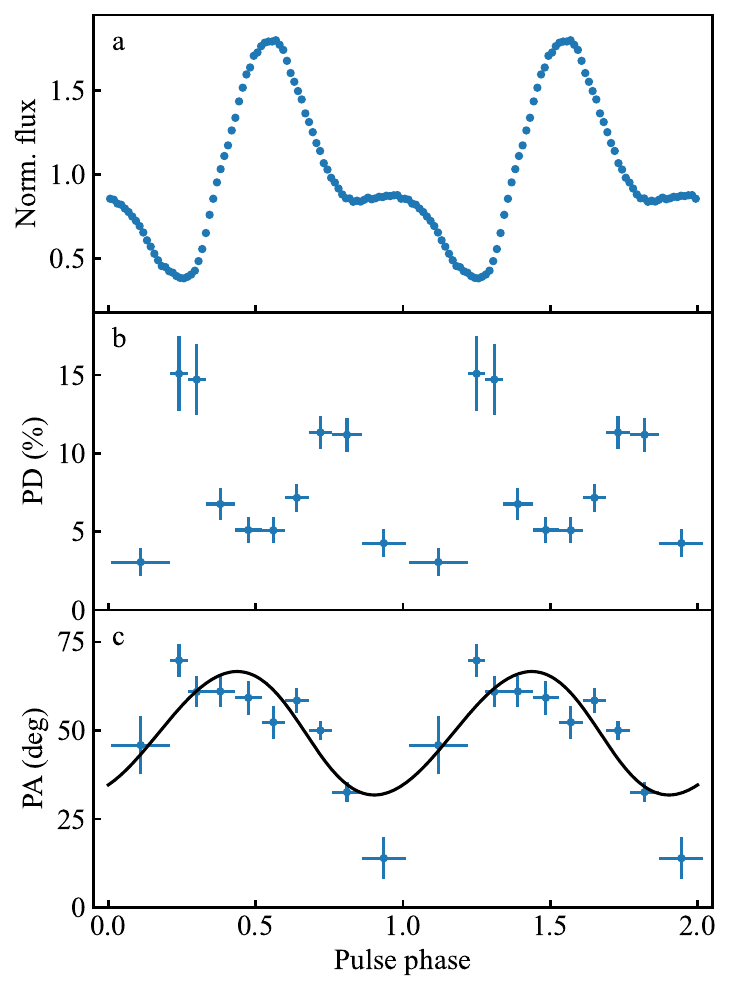}
\includegraphics[width=8.5cm]{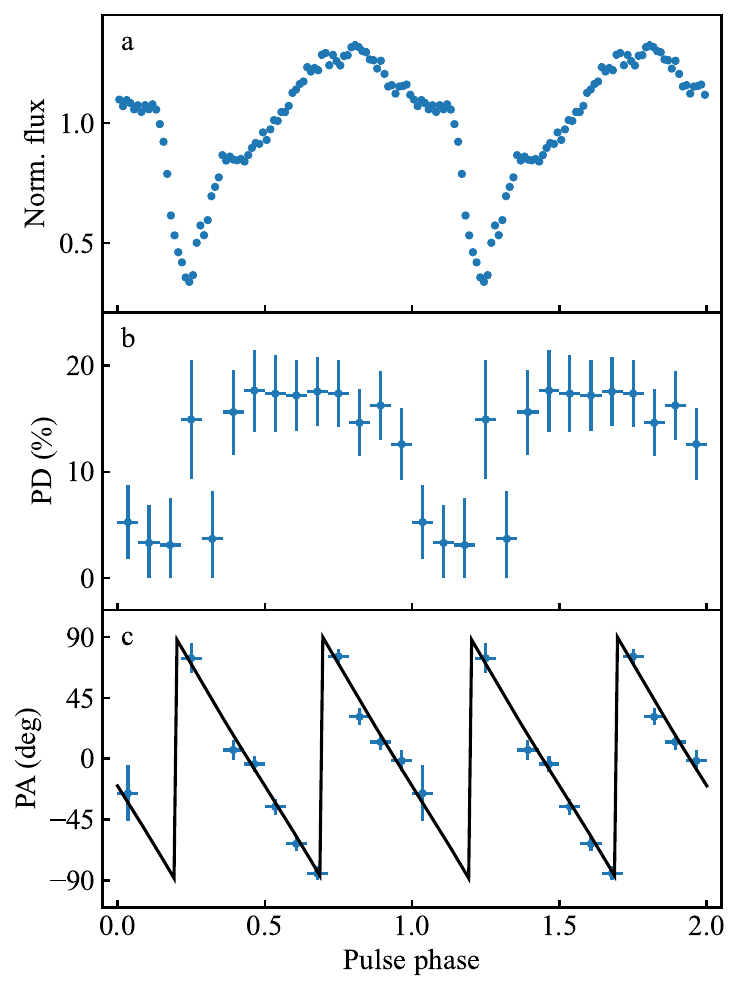} 
\includegraphics[width=8.4cm]{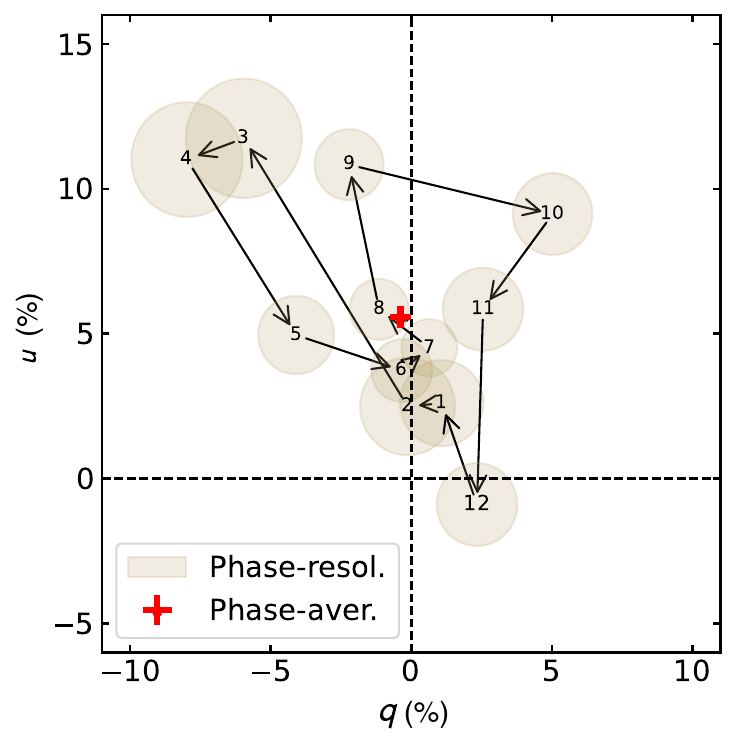}
\includegraphics[width=8.7cm]{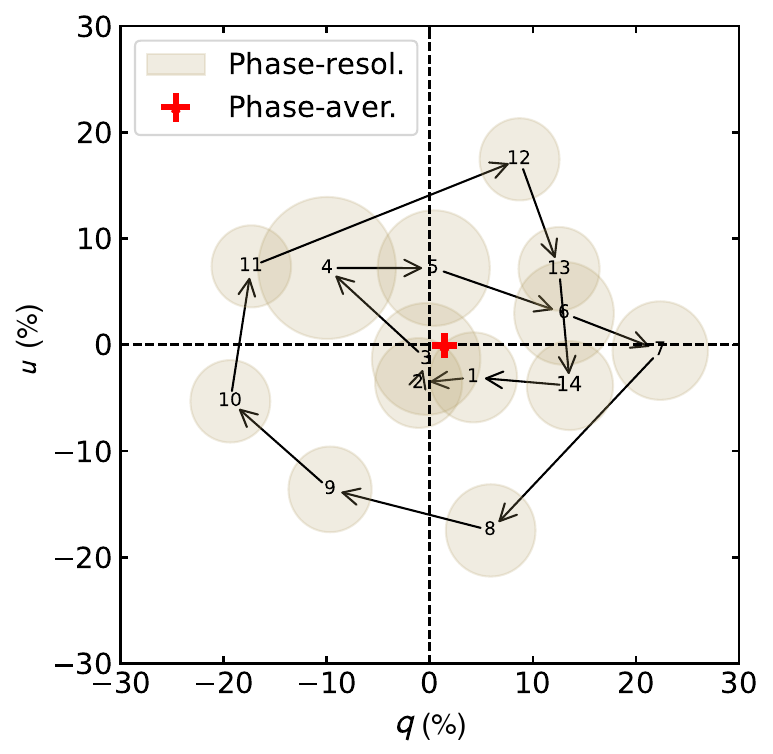}
\end{adjustwidth}
\caption{Examples of the phase-resolved spectropolarimetric analysis for two XRPs: \mbox{Cen X-3} (\textbf{left}) and \mbox{X Persei} (\textbf{right}). 
The evolutions of the normalized flux (pulse profile), PD, and~PA are shown in the corresponding panels (\textbf{a}--\textbf{c}). 
The lower panels demonstrate the phase-resolved behavior of the normalized Stokes parameters $q$ and $u$.
\label{fig:twoexamples}}
\end{figure}

After splitting all photons from \mbox{Cen X-3} into 12 bins over the spin cycle of the NS, significant variations in both PD and PA over the pulse phase were discovered. 
In~particular, it was found that PD shows a significant anticorrelation with the pulsed flux (inline with the predictions by~\cite{Meszaros88} for a sub-critical XRP), reaching a maximum value of around 15\%, which is still much lower than the theoretical predictions (see left panels in Figure~\ref{fig:twoexamples}). 
The~authors discuss several mechanisms potentially responsible for the low value of PD (see below). 
The~PA value was also found to vary over the pulse phase (Figure~\ref{fig:twoexamples}, left). 
As~discussed in Section~\ref{sec:rvm}, such variation can be used to derive the geometrical parameters of the NSs in XRPs. 
For~\mbox{Cen X-3}, the~pulsar spin position angle and the magnetic obliquity were estimated to be about 49\degr\ and 17\degr, respectively (see Table~\ref{tab:rvm} for RVM parameters).

\begin{table}[H] 
\caption{RVM parameters of the XRPs observed by \ixpe.}
\begin{tabularx}{\textwidth}{CCCC}
\toprule
\textbf{Name}	&\boldmath\textbf{$i_{\rm p}$ [deg]}	& \boldmath\textbf{$\theta_{\rm p}$ [deg]}	& \boldmath\textbf{$\chi_{\rm p}$ [deg]}	 \\
\midrule
Cen X-3		        & 70.2 (fixed)		& $16.4\pm1.3$		& $49.2\pm1.1$		  \\
\mbox{Her X-1	(main-on)}	        & $56^{+24}_{-20}$			& $3.7^{+2.6}_{-1.9}$	& $42\pm2$ \\
\mbox{Her X-1 (short-on)}        & $90\pm30$			& $16.3^{+3.5}_{-4.1}$			& $57.9\pm2.1$ \\
GRO J1008$-$57	    & $130\pm3$		& $74\pm2$		& $75\pm4$ \\
EXO 2030+375	    & $128^{+8}_{-6}$		& $60^{+5}_{-6}$		& $-30^{+4}_{-5}$		 \\
X Persei		    & $162\pm12$		& $90\pm15$		& $70\pm30$	\\
GX 301$-$2		    & $135\pm17$		& $43\pm12$		& $135$	 \\
\mbox{LS V +44 17}/Obs.~1		    & $56\pm12$		& $27\pm4$		& $82\pm1$		 \\
\mbox{LS V +44 17}/Obs.~2		    & $102\pm2$		& $54\pm1$		& $-6.2\pm0.4$ \\
\mbox{LS V +44 17}~\textsuperscript{a}		    & $108\pm2$		& $48\pm1$		& $-8.4\pm0.6$ \\
\mbox{Swift J0243}/Obs.~1	& $80\pm3$		& $87\pm2$		& $-70\pm4$		  \\
\mbox{Swift J0243}/Obs.~2 	& $60\pm5$		& $88\pm3$		& $-87\pm7$		 \\
\mbox{Swift J0243}/Obs.~3  	& $33\pm7$		& $75\pm5$		& $-66\pm7$		  \\
\mbox{Swift J0243}~\textsuperscript{a}  	& $25^{+8}_{-17}$		& $77^{+2}_{-29}$		& $-44^{+12}_{-13}$	  \\
SMC X-1		        & $91^{+41}_{-42}$		& $13^{+7}_{-6}$		& $87\pm4$	\\
\bottomrule
\end{tabularx}
\noindent{\footnotesize{\textsuperscript{a} Obtained using two-component model to the combined data set.}} 
\label{tab:rvm}
\end{table}



\begin{figure}[H]
\begin{adjustwidth}{-\extralength}{0cm}
\includegraphics[height=8.3cm]{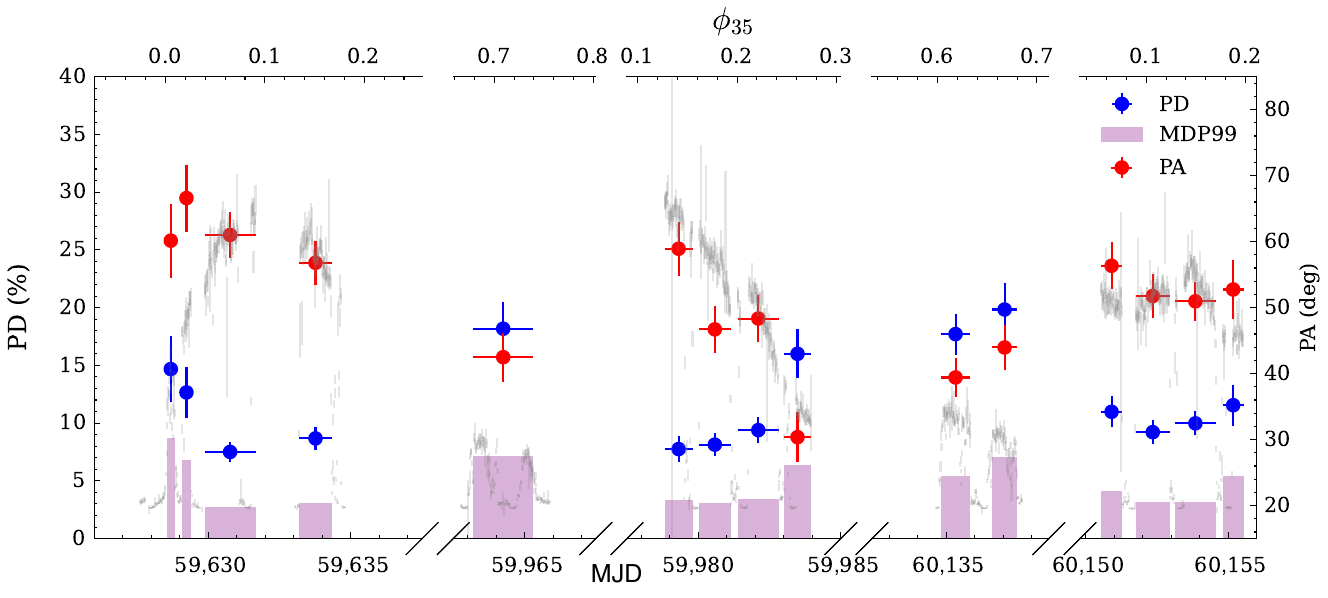}
\end{adjustwidth}
\caption{ Evolution of the pulse-phase average PD (blue circles, left axis) and the PA (red circles, right axis) of \mbox{Her X-1} with time (lower axis) and the super-orbital phase (upper axis). 
The pink rectangles show the minimum detectable polarization at 99\% CL. 
The gray symbols show the flux evolution.  
From~\cite{Zhao2024}.
\label{fig:herx1}}
\end{figure}

Another famous XRP, \mbox{Her X-1}, was observed several times during the first two years of \ixpe. 
The~reason for that is a 35 d super-orbital periodicity presumably related to free or/and forced precession of the NS in this unique system (see \citep{kolesnikov2022} and references therein). 
In~total, \ixpe performed five observations of the source, three times in the so-called ``main-on'' state and two times during the ``short-on'' state. The first two observations were utilized to determine the geometrical parameters of the NS using the RVM. 
Moreover, combining X-ray and optical polarimetric data, it was found that the spin axis of the NS and the angular momentum of the binary orbit are misaligned by about 20\degr\  \citep{Doroshenko22}. 
In addition to that, a~marginally significant variability of PA over the super-orbital cycle was revealed  (Table~\ref{tab:rvm}). 
This effect was later confirmed with additional observations of the source (see \mbox{Figure~\ref{fig:herx1}) \citep{Heyl24,Zhao2024}. }
The detailed modeling of the pulse phase-resolved data at different super-orbital phases allowed~\cite{Heyl24} to conclude that the observed 35 d period in \mbox{Her X-1} is set by the free precession of the NS crust, implying that its shape is slightly asymmetric, by~a few parts per ten million. 
In addition to that, the~authors found indications that the NS geometry is altered by torques on a time scale of a few hundred~days.

Thanks to the energy resolution capabilities of \ixpe, it is possible to study the energy dependence of the polarization properties of XRPs, albeit in a rather narrow range of 2--8~keV. 
The majority of sources from our sample do not show any significant dependence of PD or PA on energy. 
However, for~\mbox{X Persei}, a very strong increase in the PD from $\sim$$0$\% at 2 keV to $\sim$$30$\% at 8 keV was discovered~\cite{Mushtukov23}.
The reason for such behavior is still under debate. 
Applying the RVM to the phase-resolved polarimetric data allowed the authors to demonstrate that the magnetic obliquity in this source is very close to 90\degr, making \mbox{X Persei} a so-called orthogonal rotator (see right panels of Figure~\ref{fig:twoexamples}  and compare to the red dotted line in Figure~\ref{fig:rvm}d). 
Moreover, the~direct correlation of the PD and flux over the pulse phase observed in this source contrasts with the pattern observed in \mbox{Cen X-3} (compare left and right panels in Figure~\ref{fig:twoexamples}) and the predictions~\cite{Meszaros88}. 
A possible reason for such behavior is an inverse temperature profile in the NS atmosphere at low mass accretion rates (see below) and, correspondingly, the modified beam pattern~\cite{Mushtukov23}.

In the case of \mbox{Vela X-1}, the~energy dependence of the polarimetric properties looks even more interesting. First, at~low energies, the~PD value decreases with energy, reaching a minimum around $\sim$$3.5$~keV \citep{Forsblom2023}. 
Above this energy, it starts to increase, reaching around 10\% at 8 keV. 
However, the~most fascinating fact is that the PA makes a 90\degr swing between low and high energies. 
This, combined with a very complex pulse profile shape, results in very complicated phase-resolved variations of the PA, which cannot be fit with the RVM. 
Therefore, \mbox{Vela X-1} remains an XRP with still unknown~geometry.

Several sources were observed several times as a function of their luminosity. 
One such XRP is GRO~J1008$-$57, observed by \ixpe two times during the outburst in November 2022 \citep{Tsygankov2023}. 
The luminosities of the source during both observations differed by only a factor of $\sim$2, resulting in almost identical polarimetric properties, which allowed authors to increase counting statistics by combining two data sets. 
Phase-resolved spectro-polarimetric analysis of the combined data revealed a picture very similar to that of \mbox{X Persei}: positive correlation of the PD with the pulsed flux and large magnetic obliquity ($\sim$$75$\degr).

The only extragalactic XRP, \mbox{SMC X-1}, was observed with \ixpe also several times covering a significant fraction of its $\sim$45 d super-orbital cycle \citep{Forsblom24}. 
Another distinctive feature of this pulsar is a very high observed luminosity ($L_{\rm 2-8 \,keV}\approx2\times10^{38}$\,erg\,s$^{-1}$), making \mbox{SMC X-1} the most luminous XRP ever observed by \ixpe. 
In spite of the clearly super-critical regime of accretion, the polarization properties of the source are very similar to the ones observed from much weaker XRPs \mbox{Cen X-3}: low peak PD value around 10\% and anticorrelation of the PD with the flux over the pulse phase. 
Applying the RVM to three consecutive observations allowed the authors to reveal a gradual drift of the position angle of the pulsar, possibly indicating scattering in the wind of the precessing accretion disk observed at a large~inclination.

In addition to the XRPs discussed above, \ixpe observed several objects where the counting statistics did not allow for a detailed analysis. 
Specifically, in~\mbox{GX~301$-$2} \citep{Suleimanov2023} and \mbox{EXO~2030+375} \citep{Malacaria23}, polarization was significantly detected only in a few phase bins. 
However, this did not preclude the authors from determining the geometrical parameters in these systems using the RVM (see Table~\ref{tab:rvm}).

Although in all XRPs discussed above, the observed PD was low, it was detected significantly in at least the phase-resolved data. 
The only exception is 4U~1626$-$67, the~only XRP from the \ixpe sample in an ultracompact low-mass X-ray binary, in~which even phase-resolved analysis did not result in significant measurement of the PD~\cite{Marshall22}. 
Only marginal evidence of a non-zero polarization of the power-law component at the $4.8\pm2.3\%$ level was found in the combined phase-resolved spectro-polarimetric modeling. 

As was mentioned above, one of the most striking results obtained by \ixpe for XRPs is a very low PD, significantly below the theoretical predictions. 
Different authors discussed multiple reasons for that. 
In particular,~\cite{Doroshenko22} suggested a physical reason related to the thermal structure of the NS atmosphere valid for XRPs accreting at sub-critical rates when the energy of the infalling matter is dissipated at the NS surface, heating its upper atmosphere. In~the case of a nearly critical mass accretion rate, the~emergent radiation escapes primarily perpendicular to the normal direction to the NS surface, forming a ``fan''-like beam pattern. 
A simplified calculation by~\cite{Doroshenko22} indicates the possibility to obtain a low PD value if the vacuum resonance, where the contributions of plasma and magnetized vacuum to the dielectric tensor cancel each other and mixing of the normal modes of radiation occurs, is located in the transition atmospheric layer with a sharp temperature gradient. 
Geometrical factors such as the combined emission observed simultaneously from two poles at different angles also might be a reason for the low PD values in XRPs with luminosities in a wide range. 
In the case of very bright XRPs ($\gtrsim$$10^{38}$\,erg~s$^{-1}$), a~substantial fraction of emission from the NS is intercepted by the  optically thick envelop and scattered, leading to the depolarization effect \citep{Poutanen24,Forsblom24}.



\begin{figure}[H]
\begin{adjustwidth}{-\extralength}{0cm}
\centering
\includegraphics[height=20 cm]{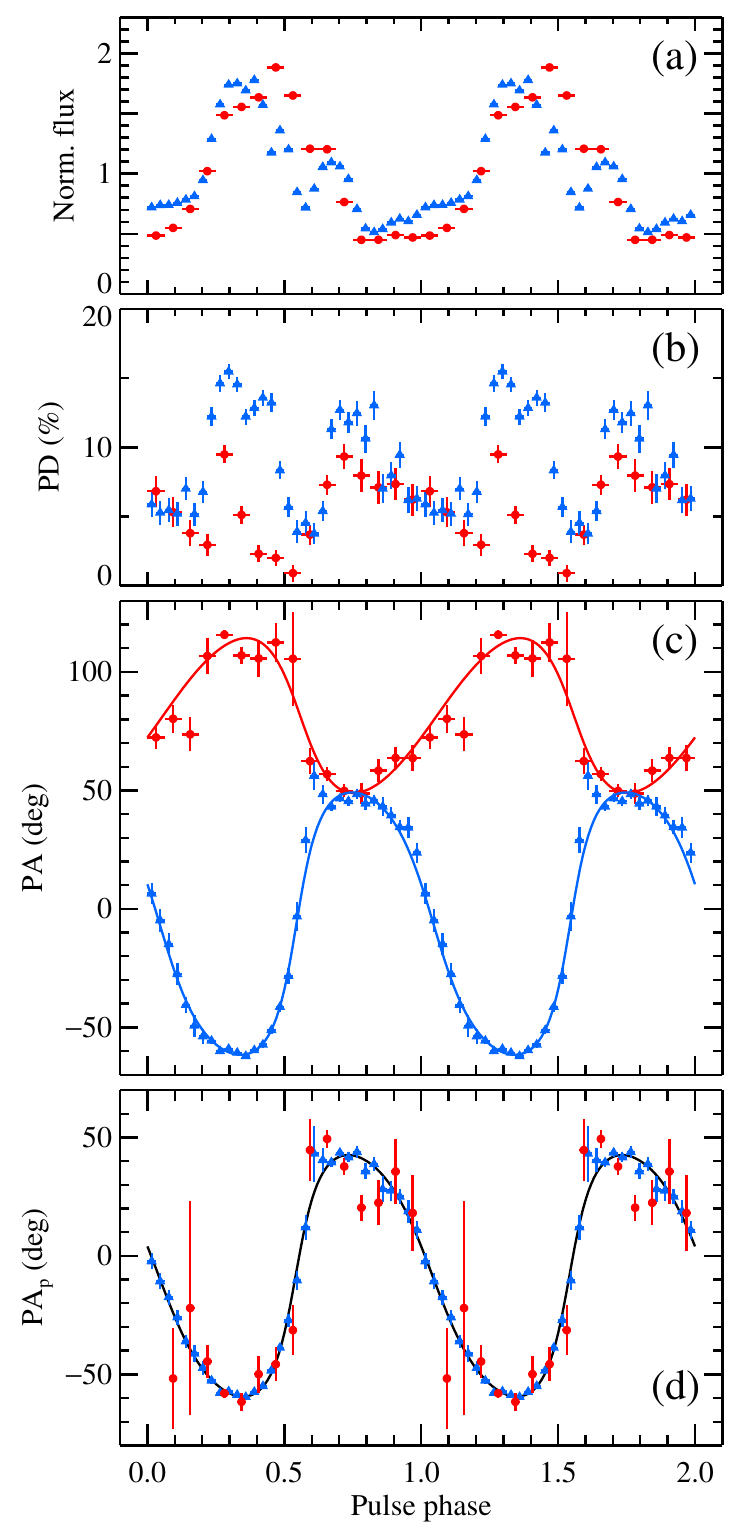}
\hspace*{0.5cm}
\includegraphics[height=20 cm]{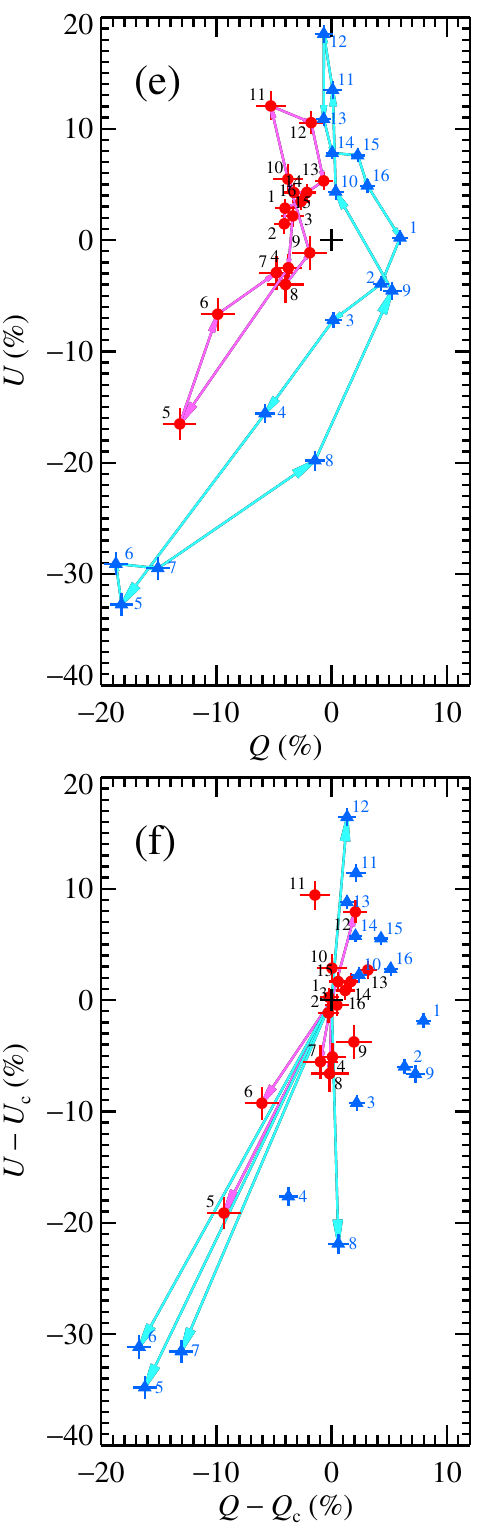}
\end{adjustwidth}
\caption{Pulse phase dependence of polarimetric characteristics of XRP \lsv. 
(\textbf{a}) Normalized flux, (\textbf{b}) PD,  (\textbf{c}) PA, and~ (\textbf{d}) intrinsic PA after subtraction of the constant component. 
Red circles and blue triangles correspond to two observations separated by two weeks. 
Adapted from~\cite{Doroshenko23}.
Panel (\textbf{e}) shows the observed phase-resolved \textit{absolute} Stokes parameters $Q$ and $U$ (scaled to the mean flux) rebinned to the same phase intervals. 
The numbers mark the bins from 1 to 16, and the arrows connect the data points. 
The black cross is the origin. 
Panel (\textbf{f}) is the same diagram with the  constant component subtracted. 
The arrows now connect the origin to a few selected points, illustrating that the arrows corresponding to the same phase bin in the two observations are nearly parallel, implying similar PA.   
\label{fig:lsv}}
\end{figure}

Most of the XRP polarimetric data can be well described by the RVM. 
However, sometimes the RVM parameters change with time. 
Their variation in \her can be explained by precession~\cite{Heyl24,Zhao2024}. 
Sometimes, the~variations are just too large for us to believe in the changes of the geometry. 
The first case of that sort was the bright transient XRP \rx/\lsv that went to the outburst in the beginning of 2023. 
\ixpe performed two observations separated by two weeks. 
The pulse profile showed some variations (Figure~\ref{fig:lsv}a), and the~PD grew from less than 10\% to more than 15\% (Figure~\ref{fig:lsv}b) when the flux dropped just by 30\%.  
To our big surprise, the~PA varied dramatically (see Figure~\ref{fig:lsv}c). 
Although the RVM fits the PA phase dependence rather well, the~parameters varied wildly (see Table~\ref{tab:rvm} and~\cite{Doroshenko23}), with, for~example, the~pulsar inclination and magnetic obliquity changed by $\sim$40\degr\ and $\sim$30\degr, respectively, and~$\chi_{\rm p}$ rotated by $\sim$90\degr. 
Although the latter change might be related to a switch between the modes, strong variations in the other parameters in just two weeks was difficult to~understand. 
A detailed look at the tracks that the source leaves on the Stokes parameter plane reveals that they are very similar  (see Figure~\ref{fig:lsv}e), but~the amplitude is larger in the second observation, and, what is even more important, the~tracks lie on the different sides of the origin, resulting in very different PAs.
This fact brought us to the idea that there are two polarized components in the \lsv data.  
The first one is associated with the pulsar and is described by the RVM. 
The second one is independent of the pulsar phase. 
In this case, the~\textit{absolute} (but possibly scaled, e.g.,~to the average flux value) Stokes parameters for each observation can be expressed as a sum of the variable and constant components:
\begin{eqnarray}  
\label{eq:two_comp}
I(\phi) &=& I_{\mathrm c} + I_{\mathrm p}(\phi) , \nonumber \\
Q(\phi) &=& Q_{\mathrm c} + P_{\mathrm p}(\phi)I_{\mathrm p}(\phi)\cos[2\chi(\phi)] , \\
U(\phi) &=& U_{\mathrm c} + P_{\mathrm p}(\phi)I_{\mathrm p}(\phi) \sin[2\chi(\phi)]  .  \nonumber
\end{eqnarray}  
Here, indices denote the constant (c) and pulsed (p) components. 
The Stokes parameters of the constant component are additional fit parameters. 
The RVM now has to fit the PA of the variable component for both observations that can be obtained from
\begin{equation}
\label{eq:chi_intr}
\chi(\phi) = \frac{1}{2} \arctan\left[ 
\frac{U(\phi)-U_{\rm c}}{Q(\phi)-Q_{\rm c}}\right] .
\end{equation} 

To illustrate this two-component model, we plot the intrinsic Stokes parameters, i.e.,~those where the best-fit constant component is subtracted, in~Figure~\ref{fig:lsv}f. 
If, before the subtraction, the~PA for the two observations at the same phases had completely different values (Figure~\ref{fig:lsv}c), then after~the subtraction, the~PAs are similar as~illustrated by the green arrows in Figure~\ref{fig:lsv}f, and~can be fitted with the common RVM (see Figure~\ref{fig:lsv}d). 

From the best-fit Stokes parameters of the constant component $Q_{\rm c}$ and $U_{\rm c}$, one can obtain the polarized flux of that component $P_{\rm c}I_{\rm c}=\sqrt{Q_{\rm c}^2+ U_{\rm c}^2}$ which turned out to be \mbox{3--4.5\%} (of the pulse-average flux) depending on the  observation. 
Interestingly, the~PA of the constant component $\chi_{\rm c}$ was found to be consistent with the PA of the optical emission $\chi_{\rm o}$ in \lsv.
These two facts were interpreted as evidence in favor of the accretion disk being responsible for the constant polarized component and the accretion disk and the Be-star decretion disk being coplanar.
Radiation scattered in an equatorial outflow is polarized parallel to the disk axis with the dependence on inclination \mbox{PD = $\sin^2i/(3-\cos^2i)$ \cite{ST85}} and thus can reach a value of 33\% for edge-on observers. 
Furthermore, the~PD drops slowly with decreasing inclination and depends rather weakly on the vertical extent of the wind. 
Alternatively, the~pulsar radiation can be scattered off the accretion curtain, where the disk interacts with the magnetosphere. 
The values of a few percent for $P_{\rm c}I_{\rm c}$ can be achieved if, for example, the constant component is polarized at a level of $\sim$20\% and it contributes $\sim$10--20\% to the total flux;  these numbers are realistic in the accretion wind scenario. 
On the other hand, the~position angle of the pulsar spin $\chi_{\rm p}$ was found to be different by $\sim$$75\degr$ (or $\sim$$15\degr$ if pulsar radiation is dominated by the X-mode) from $\chi_{\rm c}$, implying a possible misalignment between the pulsar spin and the orbital~axes. 

In the summer of 2023, ultra-luminous X-ray pulsar \swiftp went to the outburst, and \ixpe observed it three times for two days every time with about two-week intervals~\cite{Poutanen24}. 
This XRP also showed variations in polarimetric properties but not as extreme as \lsv. 
On two occasions, the~PA showed a sine-wave like pattern but with half a period and during the last observations, and the~PA showed a clear sawtooth pattern implying $i_{\rm p}<\theta_{\rm p}$. 
Because the first two data sets cannot be modeled with pure RVM, the~two-component model was applied, leading to a much better description of the data and allowing to obtain the geometrical parameters (Table~\ref{tab:rvm}). 
In this object, the~polarized flux  of the constant component $P_{\rm c}I_{\rm c}$ varied between 1.5\% and 3\%, and a $\sim$60\degr\ difference was found between $\chi_{\rm c}$ and $\chi_{\rm p}$, implying a misalignment angle between the pulsar spin  and the orbital axes exceeding 30\degr.
On the other hand, in~this object, the optical PA $\chi_{\rm o}$  and $\chi_{\rm c}$ were barely consistent with each other, indicating a possible misalignment between the accretion and decretion disk~axes.

\section{Summary}
 
All existing theoretical models of emission from accreting XRPs predict very high PD, up~to 80\%, making them the most promising targets for X-ray polarimeters. 
The launch of the \ixpe observatory provided the first opportunity to verify these predictions and thus probe our understanding of radiation transport in the presence of extremely strong magnetic~fields. 

The first two years of \ixpe operations have demonstrated a complete inconsistency between our expectations and reality. 
Specifically, in~none of the observed XRPs (around a dozen in total) did the PD exceed 15--20\%, with~typical values varying between 5 and 15\%. 
In one source, low-mass X-ray binary 4U~1626$-$67, the~polarization was not detected, with~upper limits around 5\%. 
In all other XRPs, both PD and PA showed significant variability over the pulse~phase.

In contrast to the PD, which sometimes resembles a complex pulse profile shape, the~PA demonstrates very smooth and simple variations over the pulse phase. 
This is related to quantum-electrodynamics  effects and the simple dipole configuration of the NS magnetic field at relatively large distances. 
This feature allows us to apply the RVM to independently determine the geometrical configuration of XRPs, thereby breaking the longstanding degeneracy between the geometry of the pulsar and the physics of its emission. 
Thanks to this, it was possible to confirm the free precession of the NS in \mbox{Her X-1} and discover unpulsed but strongly polarized emissions from the bright transients \lsv and \swiftp.

The growing sample of NSs with known geometry will play an important role in studies of NS spin evolution in the context of binary systems and will allow tighter constraints on the detection of continuous gravitational waves~\cite{Piccinni22,Wette23}. 
Future observations of XRPs with \ixpe will hopefully increase the statistics, allowing us to study the energy dependence of polarimetric characteristics and to shed light on the physical nature of various spectral components.
We also hope to catch an outburst of a bright transient XRP to study changes in the polarization between super- and sub-critical regimes of accretion, to learn about the dependence of the emission region structure on the mass accretion rate, and~to verify the whole paradigm of accretion onto the highly magnetized~NSs. 

\vspace{6pt}

\authorcontributions{Conceptualization and writing the review and editing, J.P. and S.S.T.; data curation, S.V.F. All authors have read and agreed to the published version of the~manuscript.}

\funding{This research was supported by the Academy of Finland grants 333112 and~349144. }

\dataavailability{\ixpe data are available at the HEASARC \url{https://heasarc.gsfc.nasa.gov/docs/archive.html}. } 

\acknowledgments{The authors are grateful to the members of the \ixpe topical working group on Accreting Neutron Stars and especially Victor Doroshenko and Jeremy Heyl for their important contribution to the analysis and interpretation of the \ixpe data on X-ray~pulsars.}

\conflictsofinterest{The authors declare no conflicts of~interest. }

\abbreviations{Abbreviations}{
The following abbreviations are used in this manuscript:\\

\noindent 
\begin{tabular}{@{}ll}
\ixpe & Imaging X-ray Polarimetry Explorer \\
XRP & X-ray pulsar \\
NS & Neutron star \\
PD & Polarization degree \\
PA & Polarization angle \\
RVM & Rotating vector model \\
CRSF & Cyclotron resonance spectral feature \\
\end{tabular}
}

\begin{adjustwidth}{-\extralength}{0cm}

\reftitle{References}


\PublishersNote{}
\end{adjustwidth}
\end{document}